# Signature of bi-modal fission in Uranium nuclei

Alok Chakrabarti, Siddhartha Dechoudhury, Debasis Bhowmick, Vaishali Naik

Radioactive Ion Beam Group,
Variable Energy Cyclotron Centre, Department of Atomic Energy,
&
Homi Bhabha National Institute -Kolkata,
1/AF, Bidhan Nagar, Kolkata-700064, India

**Abstract**

We report here the signature of bi-modal fission, one asymmetric and the other symmetric, in Uranium nuclei in the mass range $A$ = 230 to 236. The finding is unexpected and striking and is based on a model independent analysis of experimental mass distributions (cumulative yields) at various excitations from about 23 to 66 MeV in the $\alpha$ induced fission of $^{232}$Th. It has been found that the observed asymmetry in the mass distributions and the unusually narrow peak in the symmetry region, can both be explained in a consistent manner if one assumes: a) multi-chance fission, b) bi-modal fission at lower excitations (9 < E* < 25 MeV) for all the Uranium nuclei in the range $A$ = 230 to 236, and c) that the shell effects get washed out completely beyond about 25 MeV of excitation resulting in symmetric fission. The analysis has allowed a quantitative estimation of the percentages of the asymmetric and the symmetric component in the bi-modal fission. It has been found that the bi-modal fission in Uranium nuclei is predominantly asymmetric (~ 85%), which contributes in a major way to the observed asymmetric peaks, while the ~15% bi-modal symmetric fission is primarily responsible for the observed narrow symmetric peak in the mass distributions. The unusually narrow symmetry peak in the mass distributions indicates that the symmetric bi-modal fission in Uranium nuclei must have proceeded from a configuration at the bi-modal symmetric saddle that is highly deformed with a well-developed neck.

Corresponding author email: *alok@vecc.gov.in*

It is well known that the liquid drop model [1 -3] only leads to symmetric division of the fissioning nucleus into two fragments of equal mass. To explain the observed highly asymmetric mass distribution in fission of actinides, especially in thermal neutron induced fission of $^{235}$U, the liquid drop potential energy needs to be corrected for the shell and pairing effects [4]. The spherical shell corresponding to major shell closure at $N$=82 (doubly magic $^{132}$Sn) and deformed shell corresponding to $N$=88 appear to play a major role in the asymmetric split. However, it has also been found that the major shells do not always decide the split at low excitations; relatively small microscopic effects could also play an important role in the mass split [5]. In either case, the shell effect is expected to play a significant role only at low excitations of the compound nucleus and one expects a gradual washing out of the shell effects with increasing excitation [6-8] resulting in a shift from asymmetric to a completely symmetric split and consequently in symmetric mass distribution. However, the excitation energy beyond which shell effects are completely washed out is still not fully established experimentally and despite spectacular advances the development of a theoretical framework that can explain appropriately all the observed features of the nuclear fission process remains a challenging task.

As the excitation energy increases, the probability of multi-chance fission [9] increases. In multi-chance fission (MCF), fission occurs from the original compound nucleus as well as from residual nuclei left after one or more neutron evaporation. Thus first chance fission means fission of the original compound nucleus e.g., $^{236}$U for α + $^{232}$Th reaction. The second chance fission for the same projectile-target system means fission of $^{235}$U after one neutron evaporates out of $^{236}$U; third chance means fission of $^{234}$U after consecutive evaporations of two neutrons, and so on till the excitation energy of the compound nucleus after successive emission of neutrons falls below the fission barrier. Since neutron evaporation takes away on an average excitation energy that exceeds binding energy of the neutron by about 1.9 MeV, the higher chance fissions proceed increasingly from lower excitations of the compound systems. Depending upon the initial excitation, after evaporation of a few neutrons the excitation of the compound system would come down enough for shell effects to play an important role leading to an asymmetric split. Thus, because of MCF components of both symmetric and asymmetric fission, although from different fissioning nucleus, contribute to the observed fission product mass distribution at a given excitation energy.

It was found, in a number of earlier experimental studies [10 - 12] measuring cumulative yields in α + $^{232}$Th reaction at relatively higher excitations (~ 23 to 66 MeV) of the compound nucleus ($^{236}$U), that mass distribution remains asymmetric up to the highest excitation energy studied (66 MeV). It was also found that all the mass distributions in the excitation range 23 to 66 MeV contain a third peak, unusually narrow, at the symmetry region that becomes more prominent at higher excitation energies. The survival of asymmetric mass distribution at higher excitations of the compound nucleus up to about 60 MeV has also been reconfirmed in a recent on-line study [13] in which the fission fragment mass distribution (FFMD) of a number of actinide nuclei: $^{237-240}$U, $^{239-242}$Np and $^{241-244}$Pu were studied. The asymmetry and the overall nature of the mass distribution (resulting say from the fission of $^{240}$U) observed in this study [13] match quite well with reported mass distributions [10 - 12] measured earlier for α + $^{232}$Th reaction at similar excitation energy of the fissioning system ($^{236}$U). Further, it is interesting to note that the mass distribution data [13] of $^{240}$U fission at the excitation energies 40 to 50 MeV (Fig. 1h) and 50 to 60 MeV do contain signatures of a narrow peak in the symmetry region. It was shown [13] that the observed asymmetry in the mass distribution at higher excitations, could be explained quite well if MCF is assumed and the mass distributions are calculated using the fluctuation-dissipation model [15, 16], with MCF probabilities calculated using the GEF code [14].

Three peaked mass distributions were, however, observed in the fission of light actinides and Radium at low excitations [17 - 19]. This was interpreted as the co-existence of two separate modes of fission (bi-modal fission); one asymmetric and the other symmetric, of the same fissioning nucleus. Experimentally, it was also found that the total kinetic energy of the asymmetric fission mode is about 10 MeV higher as compared to the symmetric mode. This was another surprise (the first one being the bi-modal fission) since the symmetric fission is generally expected to result in higher total kinetic energies of the fragments (which arises out of Coulomb repulsion between the two fragments in touching configuration at scission) as compared to asymmetric split since the product of two charges (with constant sum, which in the case of fission of Ra nucleus is 88) is maximum when they are equal. It was only much later that a qualitative explanation for bi-modal fission and the observation of lower kinetic energy of the fragments for symmetric division was put forward [20] by tracing the shape evolution of the fissioning nucleus on a five dimensional potential energy surface, where apart from elongation, neck diameter and mass asymmetry, two more shape parameters - the deformations of the left and right nascent fragments were considered. The calculation shows that there are two fission paths: one asymmetric and the other symmetric. The symmetric path has a saddle point at a higher excitation (about 1.5 MeV higher than the asymmetric saddle) and corresponds to a much more elongated shape at the saddle as compared to the asymmetric path which has a lower barrier and less elongated shape at the saddle. Since the distance between the charge centres is much larger in the case of symmetric split as compared to the asymmetric one which has a rather compact configuration, the theory also offers a qualitative explanation for the experimental observation of the lower total fragment kinetic energy in the case of symmetric fission. The calculation also shows that the two modes have a well separated ridge and thus at appropriate excitation energies both the modes together would lead to a three peaked mass distribution. However, their calculations for even Uranium nuclei in the range $A = 228$ to 240 suggest only small chance for the survival of bi-modal fission all the way up to scission. The asymmetric and symmetric fission paths do still exist and the symmetric split would still correspond to a more elongated shape as compared to the asymmetric fission but as the authors have shown in the case of $^{234}$U, the ridge separating the two valleys is not high enough for Uranium nuclei to keep the two modes fully separated up to the scission point.

The existence of bi-modal fission was also concluded in the low energy proton induced fission of $^{232}$Th [21]. This was based on the experimental observation of two components in the total kinetic energy distribution of the fragments. The kinetic energy data suggested two kinds of scission configurations: a compact one and an elongated one in the same mass split around $A\sim$ 130. Assuming that total kinetic energy is due to the Coulomb repulsion energy between the two fragments in touching configuration at scission, the distance between the centres of the two fragments in the elongated configuration was estimated to be as large as 18 to19 fm [21].

In this letter we have presented our model independent analysis of fission mass distribution in α + $^{232}$Th reaction at various excitation energies of the compound nucleus $^{236}$U as reported in a number of earlier experimental studies [10-12]. Mass distribution data of $^{227}$Ac and $^{240}$U have also been analysed following the same method of analysis. Our goal was to see whether a model independent mathematical fit based analysis could explain all the mass distribution data at various excitation energies in a consistent way. To do so, one needs to assume MCF, the excitation energy cut off beyond which shell effects are washed out, and bi-modal fission in Uranium nuclei within a certain excitation energy range.

The first step towards fitting the mass distributions is the estimation of probabilities of multi-chance fission (MCF), which is done following the procedure adopted in reference 13. The excitation energies of different compound nuclei of Uranium after one or successive neutron evaporations are estimated on the basis of binding energy of the last neutron [22] and assuming an average kinetic energy of 1.9 MeV for the evaporated neutron [23, 13]. The fission modes of different Uranium nuclei are then classified as purely symmetric, bi-modal and purely asymmetric depending on the excitation energy of the compound nucleus undergoing fission. Above 25 MeV of excitation, the fission is assumed to be completely symmetric ('Pure Symmetric' mode) due to washing out of shell effects and below 9 MeV of excitation it is assumed to be purely asymmetric. In between 9 and 25 MeV excitation energy, the fission mode is assumed to be bi-modal. It may be mentioned that the choice of 25 MeV as the cut off energy for shell corrections is dictated by the need to obtain good fit to all the mass distribution data at different excitations simultaneously. The assignments of different fission modes to different Uranium nuclei are then done at all the incident α-particle energies following these criteria; as an example the assignments for incident α- particle energy of 71.4 MeV (excitation energy of 65.6 MeV) is given in Table 1.

The fitting has been done assuming double hump Gaussian corresponding to pure asymmetric mode and single hump Gaussian corresponding to pure symmetric mode. The mass centroids of the asymmetric humps and the symmetric hump are kept as fitting parameters. The FWHM width of the asymmetric Gaussian, although a fitting parameter, is kept around 15 mass units (as is observed in thermal neutron induced fission of $^{235}$U). The FWHM width in the pure symmetric case has been taken as 50 mass units for all cases. For the bi-modal fission the width of symmetric component of the bi-modal fission has been kept as a variable parameter. An additional constraint ensures that fitting doesn't alter the ratios of three modes (area under the Gaussian curves) as has been decided at the beginning on the basis of MCF probabilities.

The fitted mass distributions at different excitation energies are shown in Fig. 1a - h. For fitting of mass distributions 1a to 1g, a uniform error of 20% (± 10%) has been assumed for all mass distribution data except for 1h, for which errors are taken from the data [13]. In total eight mass distributions are fitted: six for various excitation energies of $^{236}$U; one for $^{227}$Ac and the other for $^{240}$U having excitation energy in the range 40 to 50 MeV. The last two (1g and 1h) were fission fragment mass distributions determined using in-beam techniques while the other six (1a to 1f) mass distributions for fission of $^{236}$U were determined by measuring cumulative yields. The cumulative yield measurements used the technique of off-line gamma spectroscopy where one measures the beta-delayed gamma rays from beta decay of post-neutron evaporation fragments. It may be mentioned that the gross features (asymmetric/symmetric/bi-modal, etc.) of the configuration at scission are not expected to be masked in the case of mass distribution determined through cumulative yield measurement (for example, an asymmetric configuration at scission would result in an asymmetric cumulative yield distribution, etc.). Further, the off-line gamma spectroscopy (beta-delayed gamma spectroscopy) measurements allow unique determination of mass ($A$) and the yields are not affected by the presence of isomers having half-lives of µs and longer, which affect the determination of fragment yields using in-beam gamma spectroscopy technique [24, 25].

It can be seen from Fig. 1 (a-h) that all the experimental mass distributions could simultaneously be fitted quite well by following the same set of criteria. It may be mentioned that GEF code calculations do not reproduce either the asymmetry or the symmetric peak for the CN excitation range analysed in this letter. To show a typical example, GEF code calculations for the case 45.4 MeV excitation of $^{236}$U is shown in Fig. 1e. It is found that to fit the mass distributions in the fission of $^{236}$U at different excitation energies, one requires bi-modal fission with about 85% asymmetric component and the rest symmetric fission component. The best fitted FWHM widths for the bi-modal symmetric fission lie in between 4 to 8 mass units. The best fit to the mass distribution data of $^{240}$U (Fig. 1h) also need a bi-modal symmetric component of ~15% and a FWHM width of about 6 mass units for the symmetric peak. The fit to the mass distribution of $^{227}$Ac assumes 100% bi-modal component (excitation energy 7 to 13 MeV) and requires a higher (as compared to Uranium nuclei) bi-modal symmetric fraction of about 25% and FWHM width of about 12 mass units. This width is much narrower as compared to FWHM (~ 50 mass units) of standard symmetric fission but is wider by about a factor of 2 as compared to FWHMs for bi-modal symmetric fissions of Uranium nuclei.

The unusually narrow widths (FWHM ~ 6 mass units) of the bi-modal symmetric fission in Uranium nuclei require a qualitative explanation / justification. Such a narrow width seems to suggest that the bi-modal symmetric fission in Uranium nuclei proceeds from highly deformed saddle configurations with well-developed necks. A well-developed neck, once formed at the saddle, would severely constrain the mass flow between the nascent fragments during the passage from saddle to scission, irrespective of whether the symmetric valley merges with the asymmetric valley or not. From this viewpoint, the earlier calculations [20] that predicted highly deformed configurations for (*A* = even) Uranium nuclei at the bi-modal symmetric saddle tend to lend support to the observation of narrow symmetric peak in the mass distributions. In the case of a deformed configuration with well-developed neck at the saddle, the mass flow between the nascent fragments would be restricted to the total number of nucleons in the volume of the neck and FWHM of the symmetric mass split is expected to be a fraction of that. A rough estimate of the FWHM is possible by assuming a cylindrical neck at the saddle. Following earlier experimental observations [21], it is reasonable to assume a neck length of about 6 fm. If one assumes further a neck diameter of about 4 fm, the cylindrical neck volume would then contain about 10 nucleons. At scission, the probability of the neck to get severed just at the middle position along the length of the neck is expected to be the maximum, which would result in a split into two equal masses. The neck can also get severed at any position along its length between the middle position and the two ends of the neck. The mass distribution in this picture would have a total width of roughly 10 mass units and this offers an explanation for the observed narrow (FWHM varying from 4 to 8 mass units) peaks in the symmetry region of the mass distributions. The actual shape at the saddle and its dynamical evolution leading ultimately to the splitting into two parts at the scission would surely be much more complex and different but this simple picture offers an explanation for the experimentally observed narrow symmetric peaks and leads to the important conclusion that bi-modal symmetric fission must have proceeded from a saddle configuration that is highly deformed with well-developed neck.

The narrow width of the bi-modal symmetric fission in Uranium nuclei also explains why in the mass distribution studied using in-beam particle identification techniques, for example the studies reported in references 26 and13, the symmetric peak is either not seen [26] or does not show up as distinctly [13] as in off-line gamma spectroscopic techniques. The mass (*A*) resolution in in-beam techniques determining FFMD is usually equal to or poorer than 4 mass units, which is not enough for a clear identification of narrow symmetric peak having FWHM of about 6 mass units. In the case of $^{227}$Ac, however, the fitted FWHM mass distribution width is much higher (12 mass units), allowing the symmetric peak to be detected in in-beam experiments. It may be mentioned that exact atomic number and mass number determination are indeed possible in in-beam fission studies using inverse kinematics and fragment separators but such studies [27] are so far limited to fragment yield determination in electromagnetically induced fission of $^{238}$U having mean excitation less than 15 MeV.

To conclude, it can be said that the assumption of bi-modal fission in Uranium nuclei with the symmetric bi-modal fission proceeding from a highly elongated configuration at saddle is necessary to fit the observed narrow symmetric peak in the mass distributions in the fission of $^{236}$U, studied at excitations ranging from 23 to 66 MeV using the α + $^{232}$Th reaction. The wider FWHM of the symmetric peak in the case of $^{227}$Ac seems to suggest that the bi-modal symmetric saddle configuration for $^{227}$Ac is less deformed as compared to the same for the Uranium nuclei. Further, to explain the survival of asymmetry up to the highest excitation (66 MeV), it is necessary to invoke MCF and to assume that the shell effects get washed out at around 25 MeV of excitation. It is also shown that these assumptions and the same fitting procedure explain equally well the FFMD data for fissions of $^{240}$U. It seems that bi-modal fission is quite a general feature of a number of Uranium nuclei.

**Acknowledgement:**

The communicating author is grateful to Department of Atomic Energy for the award of the Raja Ramanna Fellowship that has helped in carrying out this work.

Table1. Assignments of fission modes to different fissioning nuclei in α induced fission of $^{232}$Th at incident α -particle energy of 71.4 MeV. The fission probabilities for MCF are calculated using the GEF code [14].

| Eα (MeV) | GEF code | | Excitation Energy of Fissioning Nucleus (MeV) | Fission Mode |
|---|---|---|---|---|
| | Fissioning Nucleus | Fission Probability | | |
| 71.4 | $^{236}$U$_{92}$ | 0.011 | 65.6 | Pure Symmetric |
| | $^{235}$U$_{92}$ | 0.028 | 57.04 | Pure Symmetric |
| | $^{234}$U$_{92}$ | 0.064 | 49.84 | Pure Symmetric |
| | $^{233}$U$_{92}$ | 0.143 | 40.68 | Pure Symmetric |
| | $^{232}$U$_{92}$ | 0.266 | 33.34 | Pure Symmetric |
| | $^{231}$U$_{92}$ | 0.329 | 24.14 | Bi-Modal |
| | $^{230}$U$_{92}$ | 0.1414 | 16.38 | Bi-Modal |
| | $^{229}$U$_{92}$ | 0.009 | 7 | Pure Asymmetric |

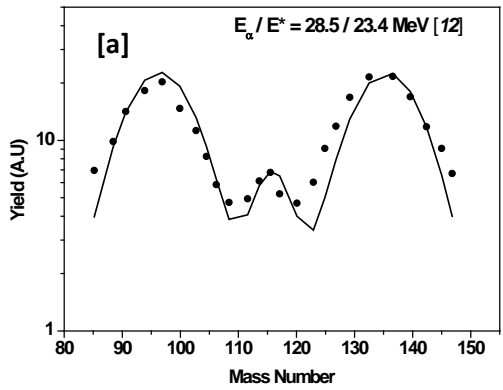
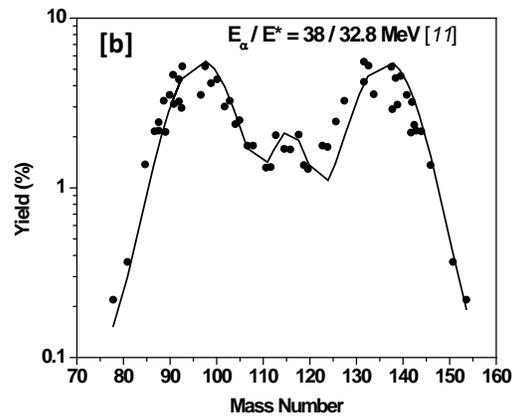
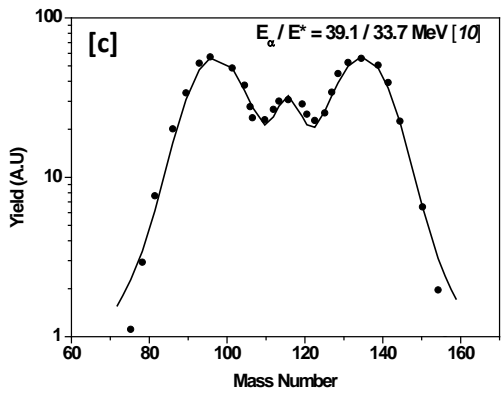
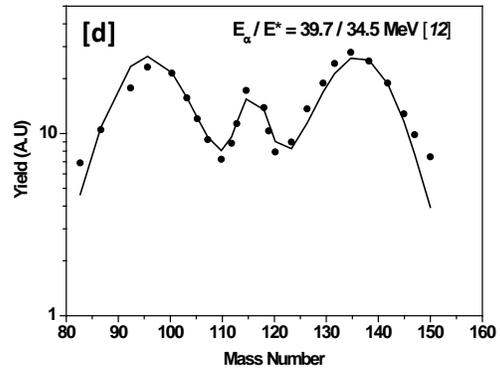
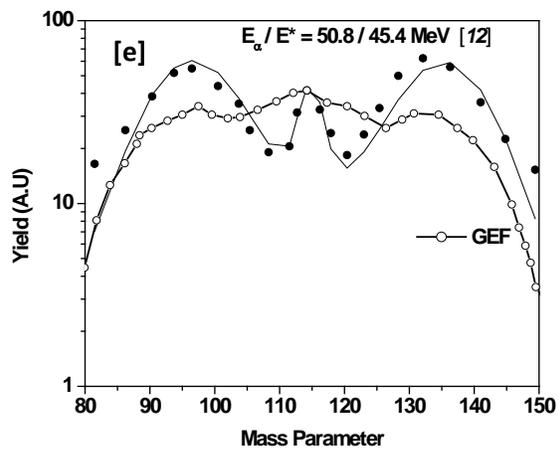
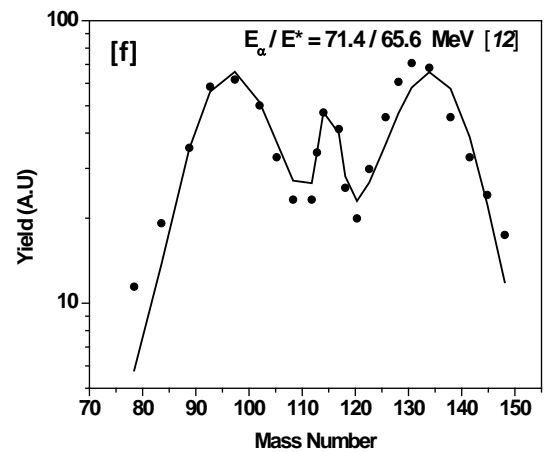

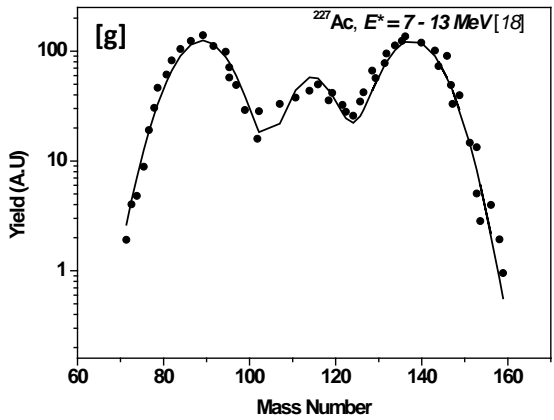 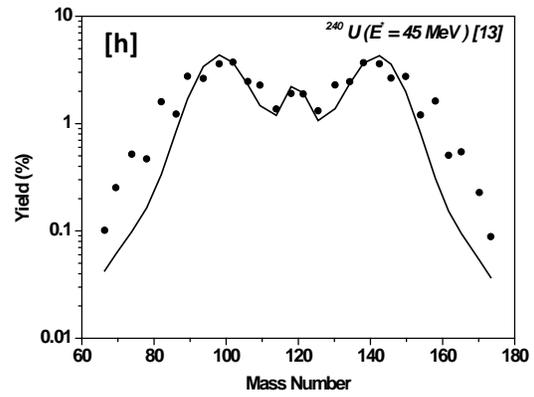

Fig.1 (a) – (f): Fitting of mass distribution data in α induced fission of $^{232}$Th at different α particle energies/excitations. The filled circles are experimental data based on measured mass distibutions taken from references indicated in the inset. The errors in the experimental data are taken as 20% (±10%) but not shown to avoid clumsiness.

Fig. 1 (g): Fitting of mass distribution of $^{227}$Ac (same 20% error in the data assumed)

Fig. 1 (h): Fitting of mass distribution of $^{240}$U at excitation energies 40-50MeV (data errors are taken from ref. 13)